\documentclass[aps,prb,showpacs,amsmath,twocolumn,amssymb,superscriptaddress,letterpaper]{revtex4}
\usepackage{times}
\usepackage{amsfonts}
\usepackage{mathrsfs}
\usepackage{graphicx}
\usepackage{dcolumn}
\usepackage{bm}
\usepackage{color}

\usepackage[colorlinks,bookmarks=false,citecolor=blue,linkcolor=red,urlcolor=blue]{hyperref}
\def\be{\begin{equation}}       \def\ee{\end{equation}}
\def\bea{\begin{eqnarray}}      \def\eea{\end{eqnarray}}

\begin{document}
\title{Staggered FeSe Monolayer on SrTiO$_3$ (110) surface}

\author{Xianxin Wu}
\affiliation{ Institute of Physics, Chinese Academy of Sciences,
Beijing 100190, China}

\author{Xia Dai}
\affiliation{ Institute of Physics, Chinese Academy of Sciences,
Beijing 100190, China}

\author{Yi Liang}
\affiliation{ Institute of Physics, Chinese Academy of Sciences,
Beijing 100190, China}
\author{Congcong Le}
\affiliation{ Institute of Physics, Chinese Academy of Sciences,
Beijing 100190, China}

\author{Heng Fan }   \affiliation{ Institute of Physics, Chinese Academy of Sciences,
Beijing 100190, China}
\affiliation{Collaborative Innovation Center of Quantum Matter, Beijing, China}

\author{Jiangping Hu  } \email{jphu@iphy.ac.cn}  \affiliation{
Institute of Physics, Chinese Academy of Sciences, Beijing 100190,
China}
\affiliation{Collaborative Innovation Center of Quantum Matter, Beijing, China}
\affiliation{Department of Physics, Purdue University, West
Lafayette, Indiana 47907, USA}

\date{\today}

\begin{abstract}
We investigate the electronic and magnetic properties of  FeSe monolayer on the anisotropic SrTiO$_3$ (110) surface. With compressive strain along $[1\bar{1}0]$ direction from the substrate,  the monolayer FeSe possesses a staggered bipartite iron lattice   with an height difference around 0.06\AA~ along the out-plane direction. The staggering causes stronger magnetic frustration between the collinear and checkerboard antiferromagnetic orders,  and the strain elongates one electron and  two hole pockets  along the strain direction   and  the remaining hole pocket  along the orthogonal direction. The strain-induced band splitting at $\Gamma$ can also result in a band inversion to drive the system into a topologically nontrivial phase. The absence of strong superconducting suppression on  the staggered lattice suggests that the superconducting pairings may be insensitive to the modification of  interactions and hopping parameters between two Fe sublattices.
\end{abstract}

\pacs{74.70.Xa, 71.18.+y, 73.90.+f}

\maketitle

\section{introduction}

Since the discovery of iron based superconductors\cite{Kamihara2008}, considerable attention has been given to the exploration of new superconducting compounds, including iron pnictides and iron chalcogenides. The iron pnictides mainly consist of 1111 family, 122 family and 111 family, where the highest T$_c$ (over 55 K) is achieved in 1111 family\cite{Ren2008,WangCao2008}. The iron chalcogenides mainly include $\beta$-FeSe\cite{Hsu2008}, FeSe$_{1-x}$Te$_x$\cite{Fang2008,Yeh2008}, the alkali-metal-doped system A$_x$Fe$_{2-y}$Se$_2$ (A=K, Rb, Cs)\cite{Guo2008} and FeS\cite{Lai2015}. Among all iron based superconductors, FeSe possesses the simplest structure but exhibits many fantastic properties. With external pressure, the T$_c$ of bulk FeSe jumps from 8 K to 37 K\cite{Margadonna2009}. Recently, the monolayer FeSe grown by molecular beam epitaxy on SrTiO$_3$ (001) surface shows a record T$_c$ of 65 K\cite{Wang2012,LiuDF2012,He2012,Tan2012}.

Many experimental and theoretical studies have been devoted to the FeSe on STO (001) surface\cite{Liu2012,Xiang2012,Bazhirov2013,Zheng2013,Bang2013,Berlijn2014,Cao2014,Huang2014,Li2014,Lee2014,Miyata2015}. The origin of high T$_c$ in the above systems is still under debate. The electron-phonon coupling may help magnetic interactions to give an enhancement of T$_c$\cite{Lee2014}. Furthermore, the non-superconducting multilayer FeSe films can become superconduting with K deposition, which clearly indicates that electron doping is  prerequisite for achieving high T$_c$\cite{Miyata2015}.

Very recently, monolayer FeSe has been successfully grown on the SrTiO$_3$ (110) surface and the onset T$_c$ of 31.6 K and a superconducting gap of 16 meV have been reported\cite{Zhou2015,Zhang2015}. Compared with STO (001) surface, the STO (110) surface is characterized by anisotropic in-plane lattice constants, thus induces a C$_4$ breaking strain on monolayer FeSe along inplane Fe-Se direction. A large isotropic superconducting gap, close to that of FeSe on STO (001) surface, has been observed in this case. This observation goes against the common belief that high T$_c$ can only be achieved in iron based superconductors with a tetragonal lattice. Thus,  the C$_4$ breaking FeSe monolayer may shed new light to the mechanism of iron based superconductors. In order to understand the underlying physics, we need to figure out the similarities and differences between FeSe on STO (001) and (110) surfaces and the effects of  the C$_4$ breaking strain.

Furthermore, the iron-chalcogenides can also be matters with nontrivial topology. Nontrivial topology was first theoretically predicted to exist in 1UC FeSe thin films on SrTiO$_3$ substrates and later was discovered in Fe(Te,Se) systems\cite{Hao2014,Wu2014,Wang2015,Xu2015}. A robust zero-energy bound
state, which is most likely be a Majorana bound state, has been observed at the interstitial iron impurity in superconducting Fe(Te,Se)\cite{Yin2015}, which supports nontrivial topology in Fe(Te,Se) materials. The C$_4$ breaking FeSe monolayer can be an intriguing material to study this nontrivial topology.

In this paper, we investigate the electronic and magnetic structures of monolayer FeSe on STO (110) surface by performing density functional(DFT) calculations. First, we obtain the electronic structure of STO (110) surface with $3\times1$ reconstruction and find that it is insulating with a large gap, consistent with previous calculations. Then, for the isolated monolayer FeSe with compressive strain along $y$ direction, we find that the iron lattice becomes a staggered bipartite lattice with a height difference around 0.06\AA~ between two Fe sublattices after relaxation but the glide-plane symmetry is still preserved. The electron pockets and two hole pockets are found to show elongation along $y$ direction,  the direction of the strain, but the big hole pocket along $x$ direction,  the orthogonal direction.  The monolayer FeSe exhibits weak hybridization with the substrate in FeSe on STO (110) surface and all Fe states are just located in the gap of the STO (110) surface layers. A low concentration of oxygen vacancies in STO (110) surface in the top layers can induce only very small electron doping to the FeSe layers. In the strained lattice, the collinear antiferromagnetic state is still the magnetic ground state in the DFT calculation but weakens compared with the lattice without strain. The strain-induced band splitting at $\Gamma$ can result in a band inversion and drives the system into a topologically nontrivial phase. These results, combined with the experimental observation of large superconducting gaps in these materials, support that the superconducting pairing in the FeSe single layer takes place dominantly  within each sublattice, which has been proposed recently to unify the understanding of the robust $s$-wave pairing symmetry in iron-based superconductors\cite{Hu2012,Hu2015}.

The paper is organized as following. In Section~\ref{S1}, the computational methods are described. In Sec.~\ref{S2}, the electronic structures of STO (110) surface, isolated monolayer FeSe and FeSe on STO (110) surface are studied. The effects of anisotropic strain and O vacancies are investigated. In Sec.~\ref{S3}, we discuss the nontrivial topology originating from the strain and the possible superconducting mechanism. Finally, we summarize and provide the main conclusions of our paper.

\section{Structure and Methods} \label{S1}
To investigate the electronic and magnetic properties of FeSe monolayer on the SrTiO$_3$ (110) surface, we performed DFT calculations. Our DFT calculations employ the projector augmented wave (PAW)
method encoded in Vienna \emph{ab initio} simulation package(VASP)
\cite{Kresse1993,Kresse1996,Kresse1996B}, and generalized gradient approximation of Perdew-Burke-
Ernzerhof\cite{Perdew1996} for the exchange correlation functional was used. After convergence tests,
the cutoff energy of 400 eV is taken for expanding the wave
functions into plane-wave basis. In the calculation, the Brillouin
zone is sampled in the \textbf{k} space within Monkhorst-Pack
scheme\cite{MonkhorstPack}. The number of these $k$ points are
depending on the lattice: $11\times11\times1$  and $11\times11\times11$ for the isolated monolayer FeSe and cubic STO, respectively. The obtained lattice constant for STO is $a=3.943$\AA,~ which is consistent with previous calculations\cite{El-Mellouhi2011}. Compared with experiments, the Se height is greatly underestimated in the paramagnetic phase of bulk FeSe\cite{Subedi2008,Lehman2010}. However, the relaxed Se height in collinear or checkerboard antiferromagnetic state is very close to experimental value. Therefore, we carried out calculations in the checkerboard magnetic state for the strained FeSe monolayer where the lattice constants were fixed and internal atomic positions were fully relaxed.

In the experiment, both $4\times1$ and $3\times1$ reconstructions on STO (110) surface have been observed\cite{Zhou2015,Zhang2015}. Here, we choose the typical STO (110) surface with the $3\times1$ reconstruction to study the monolayer FeSe on STO (110) surface.  In the calculations, the computational cell was modeled with a symmetric slab consisting of 13 atomic layers separated by a vacuum layer of 25 \AA. To model FeSe absorbed on the STO (110) surface, two FeSe monolayers were on the top of both sides of the symmetrical slab with a vacuum layer of 13 \AA. In the relaxation, the five middle layers were fixed at their bulk positions and the remaining layers including FeSe layers were allowed to relax. Forces
 were minimized to less than 0.02 eV/\AA.~ $3\times6\times1$ and $3\times3\times1$  Monkhorst-Pack $k$ point meshes were used for STO (110) surface and FeSe absorbed on the substrate, respectively. A Gaussian smearing of 0.1 eV was used in both cases.

\begin{figure}[tb]
\centerline{\includegraphics[height=3.5 cm]{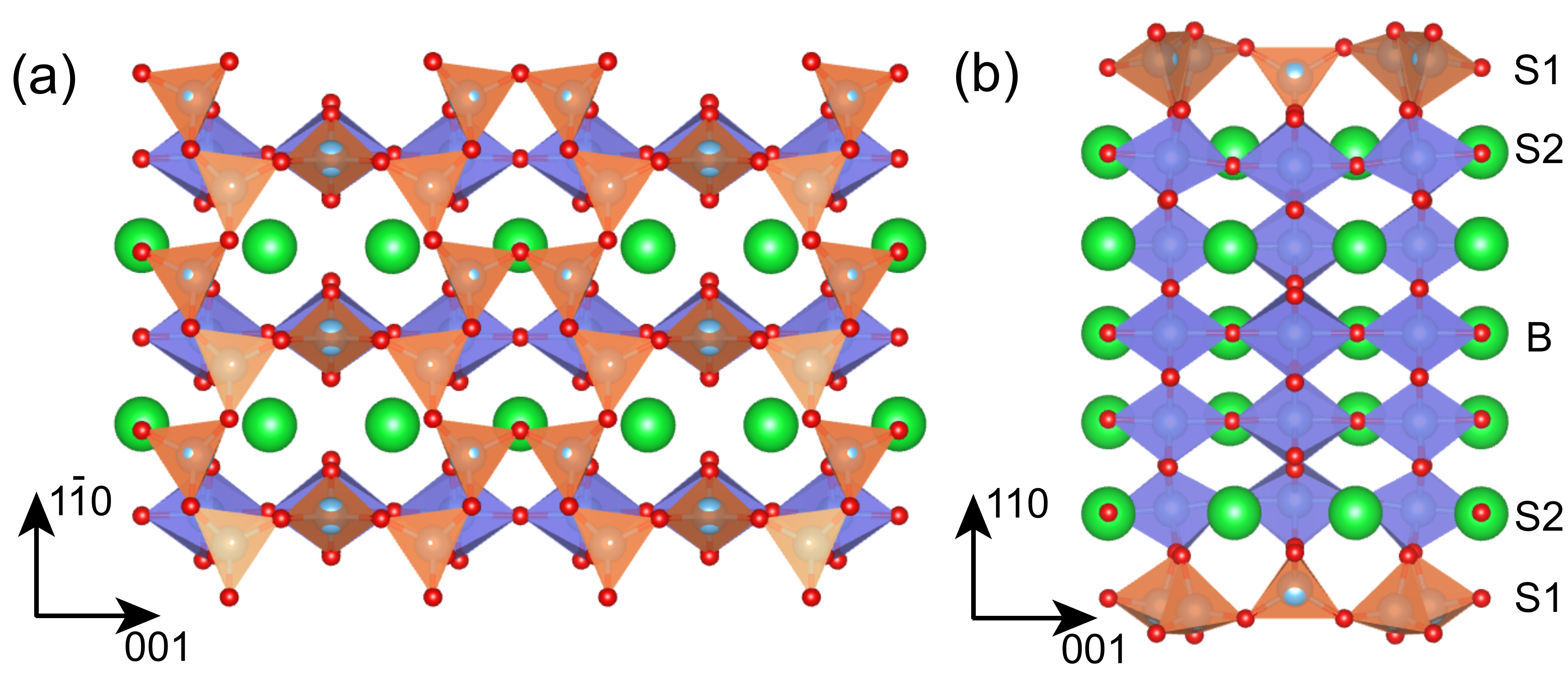}}
\caption{(color online). The surface structure of 3$\times$1 reconstruction for SrTiO$_3$ (110) surface\cite{Enterkin2010}. The TiO$_4$ tetrahedra are shown in orange and TiO$_6$ octahedra in purple. (a) top view and (b) side view of symmetric model in the calculations. The small red and big green spheres represent the O anions and Sr cations. S1, S2 and B represent for "surface1", "surface2" and bulk layers.
 \label{STO} }
\end{figure}

\section{Results and analysis} \label{S2}

\subsection{ 3$\times$1 reconstruction for SrTiO$_3$ (110) surface}
The SrTiO$_3$ (110) polar surface consists of alternating layers of SrTiO$^{4+}$ and O$_2^{4-}$, which results an unbalanced macroscopic dipole and infinite surface energy within an ionic model. Consequently, complicated surface constructions have been observed in experiment, such as $n\times1$ reconstructions\cite{Russell2008}, where $n=2,3,4,5,6$ is along the $[001]$ direction and the orthogonal direction is $[1\bar{1}0]$. The structures of $n\times1$ reconstructions have been solved with the combination of experimental and theoretical methods\cite{Enterkin2010}. The surface is composed of six or $(2n+2)$ corner sharing TiO$_4$ tetrahedra, which quenches the overall dipole moment. For example, Fig.\ref{STO} shows the surface structure of $3\times1$ reconstruction\cite{Enterkin2010}, which consists of corner-sharing TiO$_4$ tetrahedra, arranged into six- and eight-member rings. Every tetrahedron corner shares with three other surface tetrahedra except the one in the middle of eight-member ring, which is only corner sharing with two other tetrahedra. After relaxation, the atomic positions in the surface are very close to those in Ref.\onlinecite{Enterkin2010}.

\begin{figure}[tb]
\centerline{\includegraphics[height=5.8 cm]{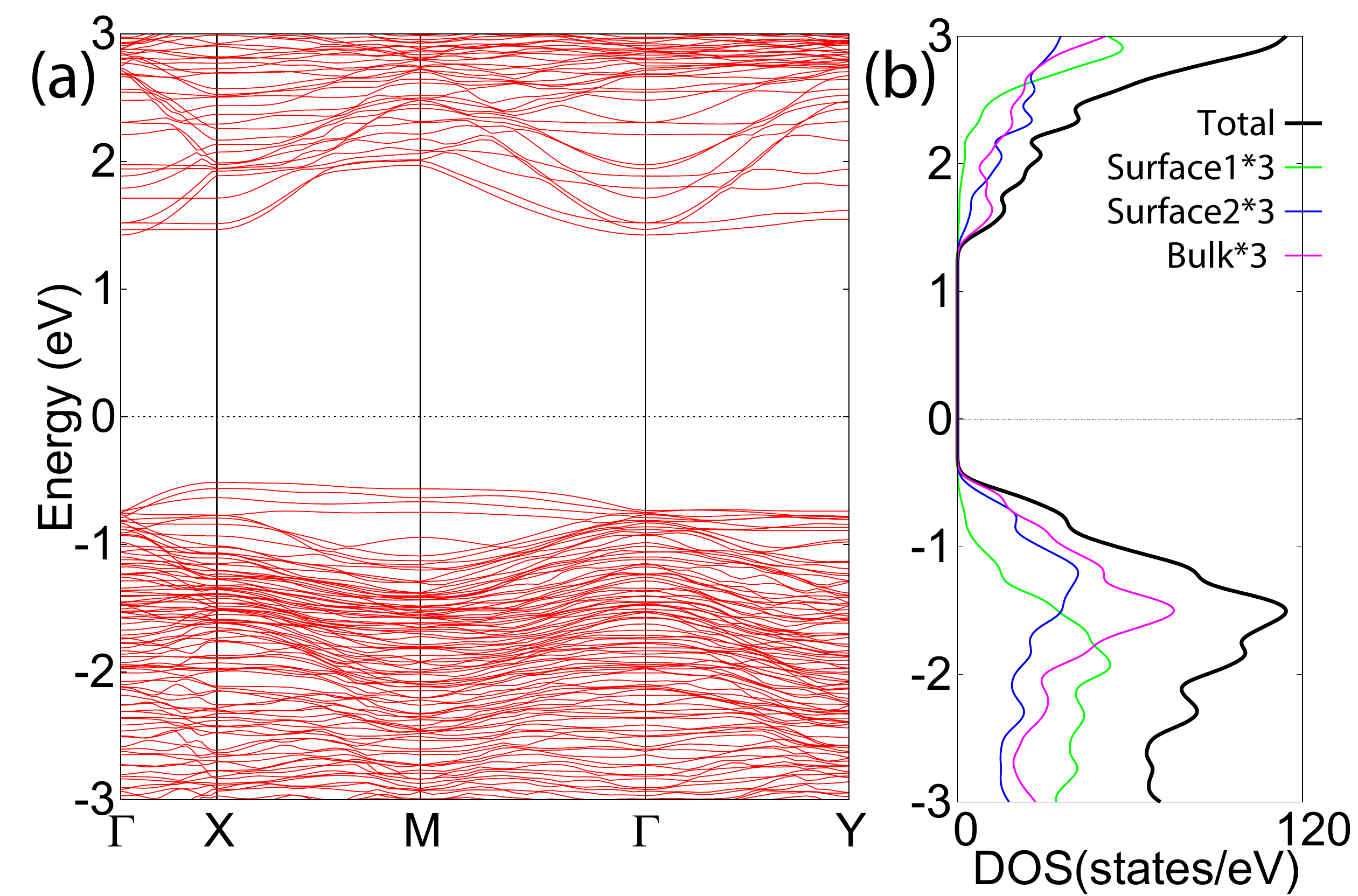}}
\caption{(color online). The band structure and density of states for the 3$\times$1 SrTiO$_3$ (110) surface. The DOS of surface atoms and bulk atoms are multiplied by a scale factor of three.
 \label{STOband} }
\end{figure}

The band structure of $3\times1$ reconstruction in a symmetric model is given in Fig.\ref{STOband}(a). It is insulating and consistent with previous calculations\cite{Enterkin2010}. The band structure is clearly anisotropic, different from that of STO (001) surface. We refer the two top or bottom layers including five Ti and thirteen O atoms as the "surface1" (S1) layer, the two layers under the S1 layer including three Ti, nine O and three Sr atoms as the "surface2" (S2) layer  and the three middle layers including fifteen O, three Ti and three Sr atoms as the bulk layer. We plot the density of states (DOS) of surface and bulk layers in Fig.\ref{STOband}(b). The gap for the bulk state is about 2.0 eV, consistent with previous results\cite{El-Mellouhi2011}, and the gap for S1 layer is much larger, about 3.0 eV, which is much larger than those of TiO$_2$-terminated and SrO-terminated STO (001) surfaces\cite{Liu2012}. The conduction bands are attributed to Ti $3d$ orbitals and the valence bands are attributed to O $2p$ states for the bulk and S2 layers.

\subsection{Monolayer FeSe with an anisotropic strain}
\begin{figure}[tb]
\centerline{\includegraphics[height=4.5 cm]{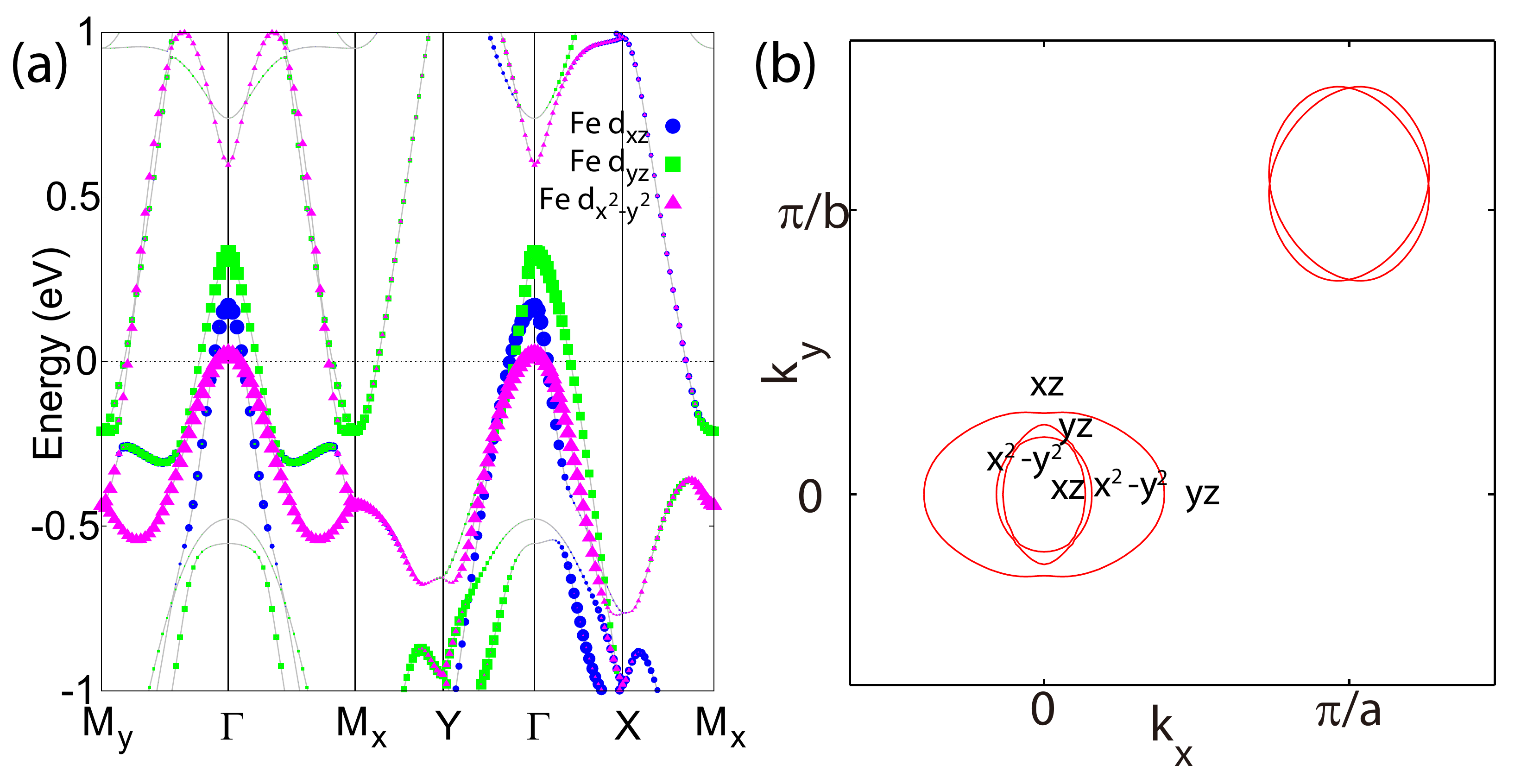}}
\caption{(color online). Band structure and Fermi surfaces for monolayer FeSe with 4\% compression along $b$ direction\cite{Zhou2015}. The special k points are : $M_x$($-\pi,\pi$), $M_y$($\pi,\pi$), $X$($\pi,0$), $Y$($0,\pi$) and $\Gamma$($0,0$).
 \label{band} }
\end{figure}

It is experimentally suggested that three unit cells of FeSe grow on the top of two unit cells of STO along $[1\bar{1}0]$ direction but an one-to-one correspondence between FeSe and STO exists along the [001] direction\cite{Zhou2015,Zhang2015}. As a result, a compressive strain is introduced along one axis ($[1\bar{1}0]$ in STO) corresponding inplane projection of the Fe-Se bonding. The difference between $a$ and $b$ of monolayer FeSe is about 4\%-5\% in experiment. To simulate the effect of $C_4$ breaking strain, we perform calculations with experimental lattice parameters $a=3.89$\AA~ and $b=3.75$\AA, corresponding to the compressive strain along $y$ direction. After full relaxation, we surprisingly find that the two Fe atoms are not in the same $z-$plane but with a height difference of 0.06\AA. The height difference persists in the relaxation with the paramagnetic state. For perfect monolayer FeSe, the point group is $D_{2d}$ at Fe site and $C_{4v}$ at Se site. However, the point group is $C_{2v}$ at both sites with such a strain. The glide-plane symmetry and inversion symmetry are still preserved in the strained lattice. The band structure of monolayer FeSe with the anisotropic strain is shown in Fig.\ref{band}(a). Because of $C_4$ breaking, the degeneracy of the $d_{xz}/d_{yz}$ bands at $\Gamma$ point is removed. The Dirac cone in $\Gamma-M$ line near the Fermi level is also gapped due to the breaking down of $C_2$ rotational symmetry around the axis of the nearest Fe-Fe bond. The glide symmetry protects the degeneracy of bands on the Brillouin zone boundary. The Fermi surfaces of strained monolayer FeSe are shown in Fig.\ref{band}(b). Both hole pockets and electron pockets are ellipses, as expected from the $C_4$ symmetry breaking. Nevertheless, there are some differences in detail between them: electron pockets and two hole pockets elongate along $y$ axis  but the other hole pocket elongates along $x$ axis. The direction of elongation for electron pockets is consistent with experiment\cite{Zhang2015}. Due to the different distributions of orbitals on hole and electron pockets, these behaviours are certainly related to orbitals. With applying compression along $y$ axis, the onsite energy of $d_{yz}$ orbital and hopping along $y$ axis increase compared with those of $d_{xz}$ orbital. Thus, the $d_{yz}$-bands shift up in energy. The orbital distribution on hole pockets can be obtained from the fat band and  is shown in Fig.\ref{band}(b). The orbital characters of the big $d_{xz/yz}$ hole pocket in $\Gamma-X$ direction and $\Gamma-Y$ direction are $d_{yz}$ and $d_{xz}$ and those of the small hole pocket reverse. With $d_{yz}$ bands shifting up , the big hole pocket consequently elongates along $x$ axis while the small one elongates along $y$ axis.  Because of the enlargement of BZ along $y$ axis, the hole pocket attributed to $d_{x^2-y^2}$ elongates along this direction. For the electron pockets, they also elongate along $y$ direction as this strain has little effect on the bands on the Brillouin boundary, which are protected by symmetry.

\begin{figure}[t]
\centerline{\includegraphics[height=6 cm]{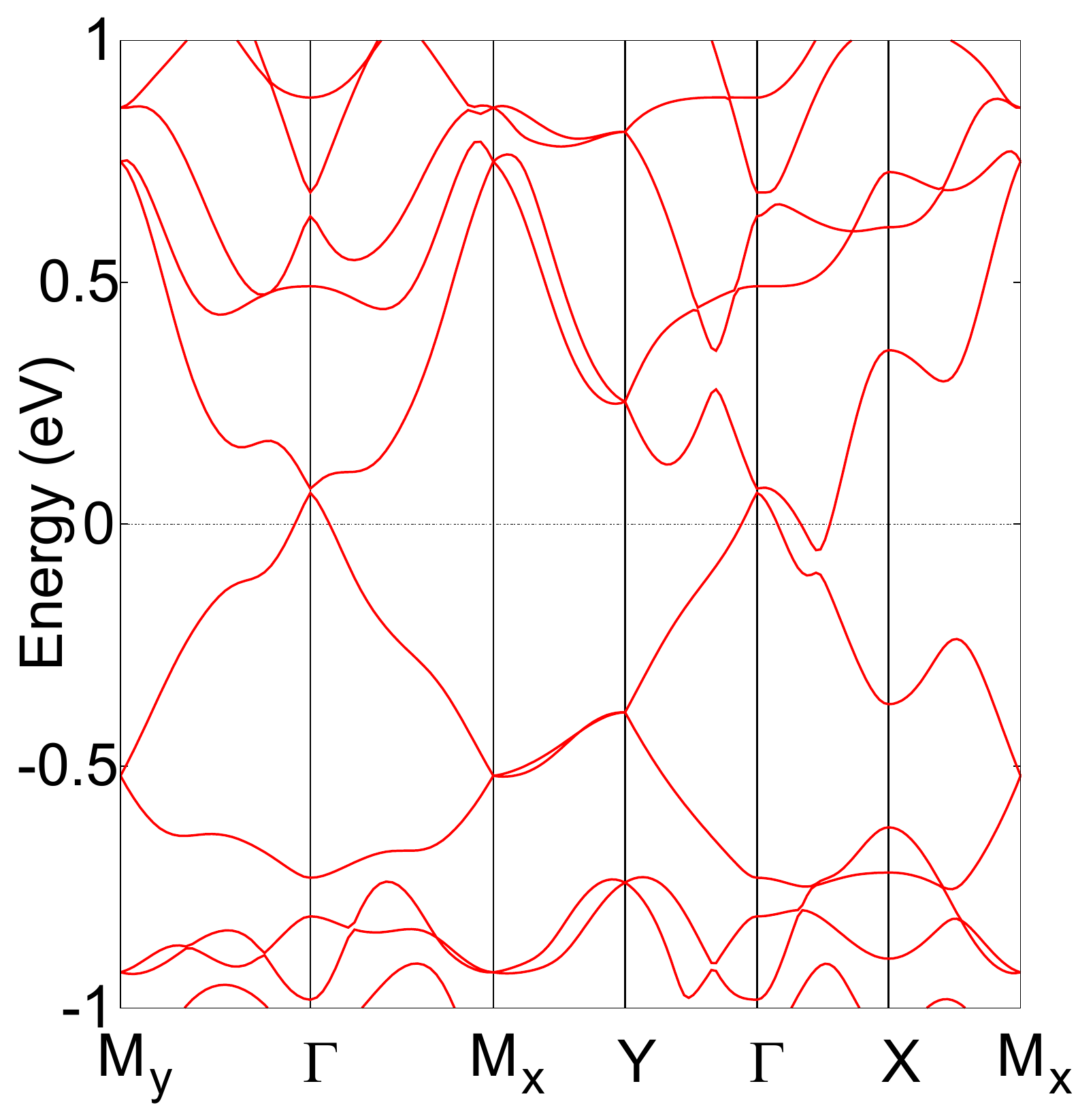}} \caption{(color online). Band structure for strained FeSe monolayer with the CAFM state. Spins are aligned antiparallel along $\Gamma X$ and parallel along $\Gamma Y$ direction. The special k-points are the same as those in Fig.\ref{band} but in small BZ of $\sqrt{2}\times\sqrt{2}$ unit cell.
 \label{edgecol} }
\end{figure}

To investigate magnetic properties for strained FeSe monolayer, we consider four magnetic states: paramagnetic (PM) state, ferromagnetic state, checkerboard antiferromagnetic (CBAFM) state and collinear AFM (CAFM) state. According to our calculations, the CAFM state is the ground state with an energy gain of 156 meV/Fe relative to PM state and a spin moment of 2.16 $\mu_B$ for each Fe. Compared with FeSe without strain, we find that the energy gains of CBAFM and CAFM states relative to PM state decrease in strained FeSe, which indicates that the $C_4$ breaking strain suppresses both AFM states. Furthermore, the energy difference between CBAFM and CAFM states decreases, which suggests that this strain induces stronger competition and magnetic frustrations between these states. The band structure of FeSe monolayer in CAFM state is shown in Fig.\ref{edgecol}. It is metallic and the $C_4$ breaking of bands from the strain and CAFM state can be clearly seen. A Dirac cone-like band appears at $\Gamma$ point.


\begin{figure}[t]
\centerline{\includegraphics[height=3.1 cm]{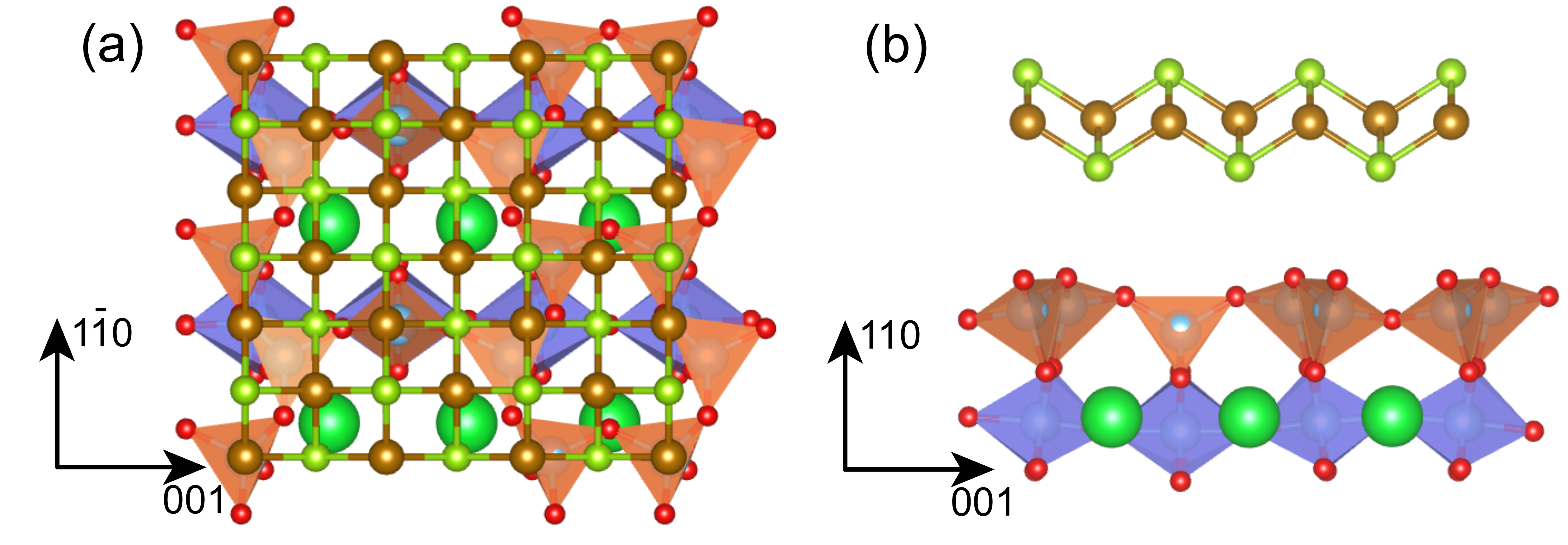}} \caption{(color online). The structure of monolayer FeSe on STO (110) surface with $3\times1$ reconstruction: (a) top view and (b) side view. The small red and big green spheres represent the O anions and Sr cations and the small brown and light green spheres represent the Fe and Se atoms.
 \label{FeSe1} }
\end{figure}

\begin{figure}[tb]
\centerline{\includegraphics[height=7 cm]{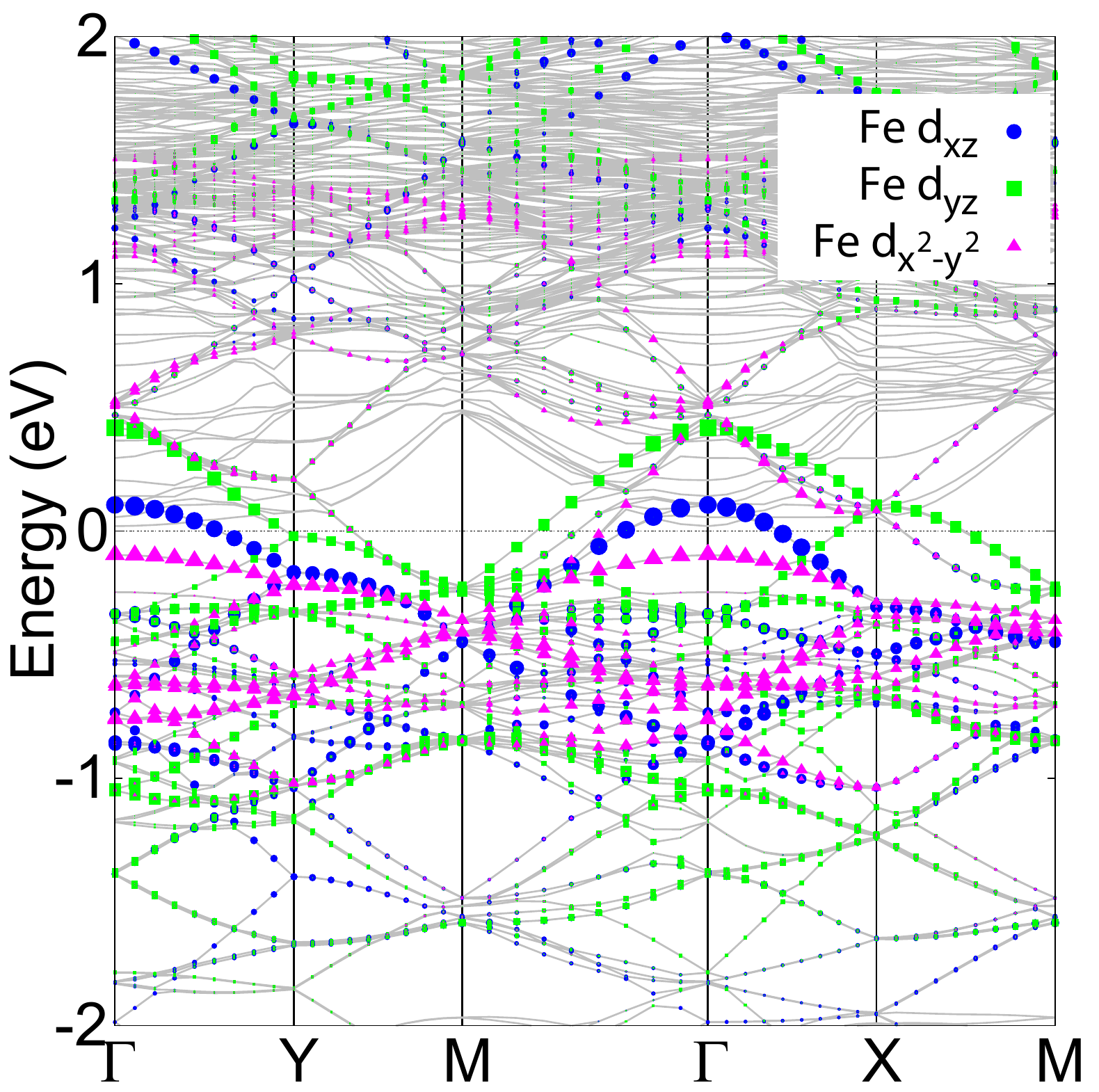}}
\caption{(color online). Band structure of monolayer FeSe on STO (110) surface ($1\times2$ unit cell for $3\times1$ reconstruction) in nonmagnetic state. The special k points are : $M$($\pi,\pi$), $X$($\pi,0$), $Y$($0,\pi$) and $\Gamma$($0,0$).
 \label{bandSTOFeSe} }
\end{figure}

\begin{figure}[t]
\centerline{\includegraphics[height=9 cm]{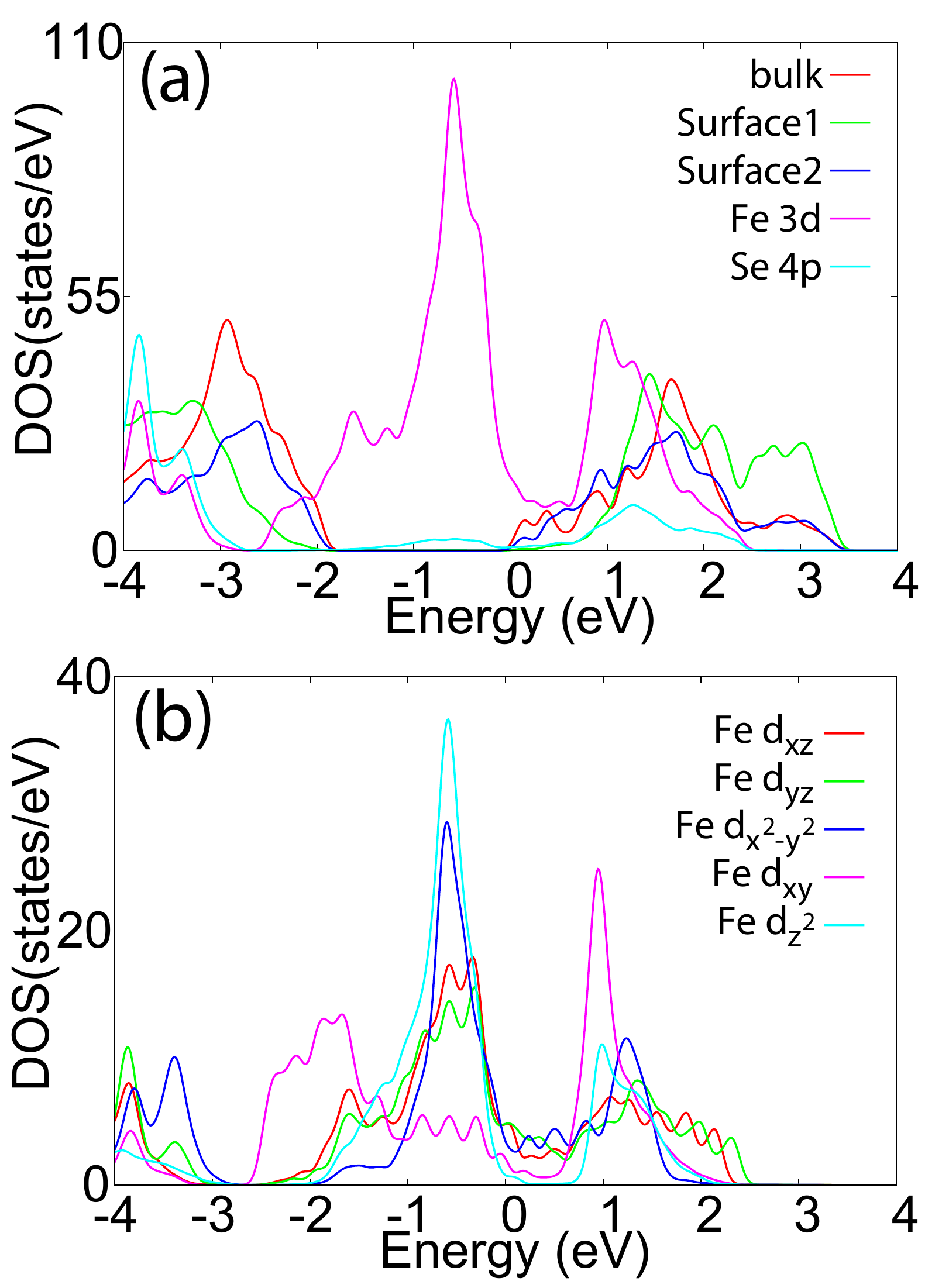}} \caption{(color online) The partial DOS of surface and bulk atoms (a) and orbit-resolved partial DOS of Fe $d$ orbitals (b) for FeSe on STO (110) surface in nonmagnetic state.
 \label{dosSTOFeSe} }
\end{figure}

\begin{figure}[tb]
\centerline{\includegraphics[height=10 cm]{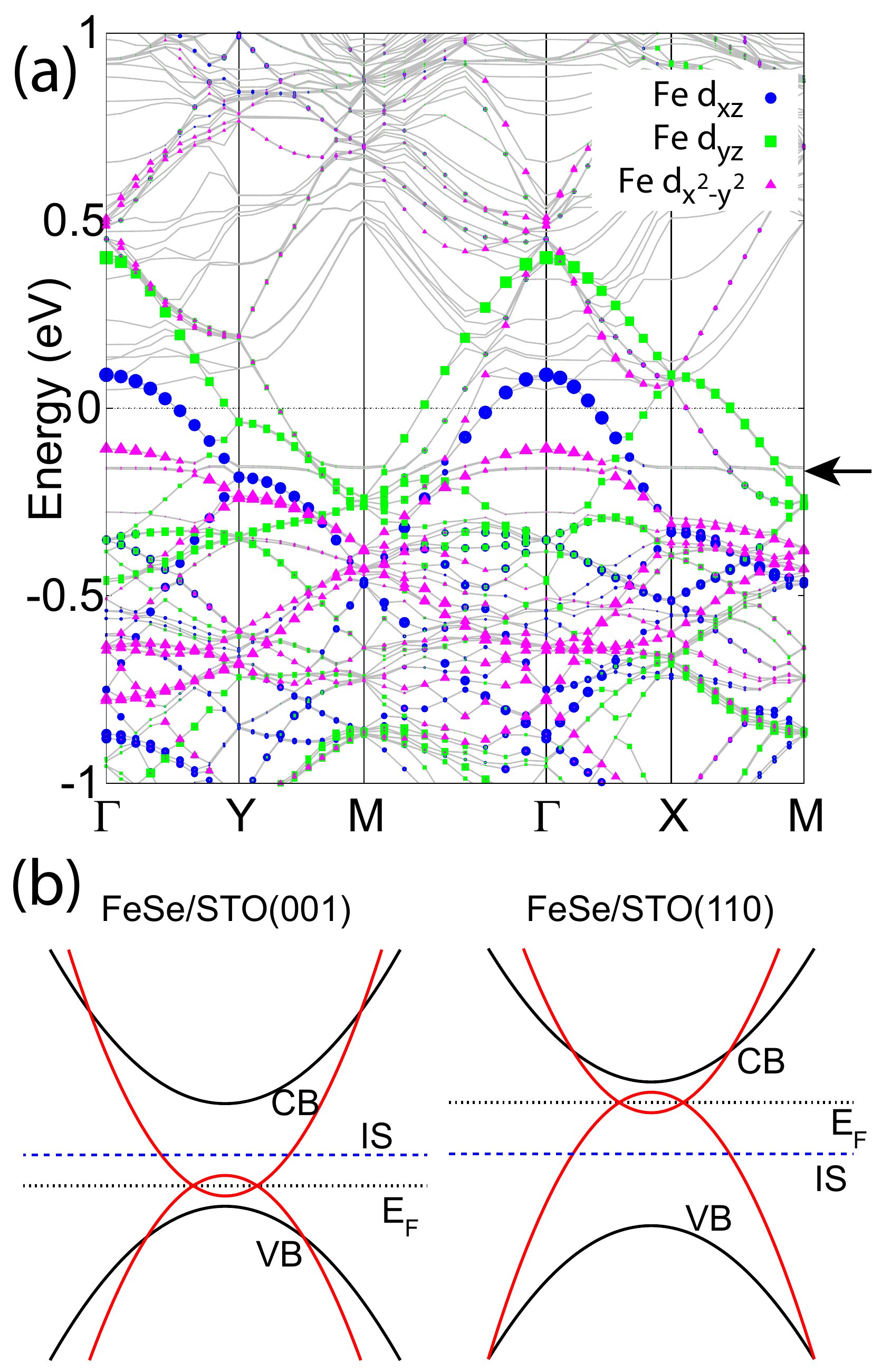}}
\caption{(color online). (a) Band structure of monolayer FeSe on STO (110) surface ($1\times2$ unit cell for $3\times1$ reconstruction) with an oxygen vacancy in the top layer in nonmagnetic state. The impurity-induced state is denoted by the black arrow. (b) Band alignments for FeSe on STO (001) surface and STO (110) surface. The solid black and red lines represent the bands of the subtrate and FeSe, respectively. The blue dashed line denotes the impurity state (IS) and the black dashed line denotes the Fermi level.
 \label{bandSTOFeSevac} }
\end{figure}

\subsection{Monolayer FeSe on SrTiO$_3$ (110) surface}

We consider monolayer FeSe absorbed on the STO (110) surface. The structure of the monolayer FeSe on STO (110) surface is shown in Fig.\ref{FeSe1}. In the calculations, we consider two possible absorption structures. One is that one Fe atom locates on the top of an TiO$_4$ tetrahedron (shown in Fig.\ref{FeSe1}) and the other one is that the bottom Se atom locates on the top of an TiO$_4$ tetrahedron (not shown). The total energy of the former case is lower by 0.2 eV per surface cell than that of the latter one. In both cases, the vertical distance of the bottom Se to the surface TiO$_4$ plane is about 3.1\AA~ after relaxation. Similar to the isolated monolayer FeSe, there is a height difference of 0.1 \AA~ between the two Fe sublattices, which is larger than that of isolated FeSe due to a larger anisotropic strain.

The electronic structures for both cases are rather similar and the band structure and DOS for latter case are shown in Fig.\ref{bandSTOFeSe} and Fig.\ref{dosSTOFeSe}, respectively. As the van der Waals interaction between FeSe and substrate is relatively weak, the band structure is just the superposition of bands for two isolated systems. The Fermi level of whole system is pinned on the bottom of the conduction band of STO (110) surface. It is in sharp contrast to that of FeSe on STO (001) surface, where the Fermi level is near the top of the valence band\cite{Liu2012}. The band alignments for FeSe on STO (001) and STO (110) surfaces are given in Fig.\ref{bandSTOFeSevac}(b). From Fig.\ref{dosSTOFeSe}(a), we find that the most of $3d$ states of Fe atoms locate in the gap of the substrate. Near the Fermi level, the states mainly attributed to Fe $3d$ orbitals of bulk and S2 atoms of the substrate, suggesting the weak coupling between FeSe and the S1 layer. The $d_{xz}$, $d_{yz}$ and $d_{x^2-y^2}$ bands at $\Gamma$ point locate at 0.105, 0.415 and -0.096 eV relative to the Fermi level. As the compressive strain is along $[1\bar{1}0]$ ($y$-axis), the $d_{yz}$ bands are pushed up compared with $d_{xz}$ band. There are Fermi surfaces around four special $k$-points due to the band-folding effect from the large unit cell. The hole pockets at $\Gamma$ and electron pockets at M are very similar to those of isolated FeSe monolayer. Near the Fermi level, $d_{xz}$, $d_{yz}$ and $d_{x^2-y^2}$ orbtials contribute dominantly, as shown in Fig.\ref{dosSTOFeSe}(b), similar to the bulk FeSe.

Next we consider the magnetic ground state of monolayer FeSe on STO (110) surface. Considering limitation of the computational resources, we only perform calculations for the system without the substrate with the four considered magnetic states in isolated FeSe. The ground state is still found to be the collinear magnetic state with an energy gain of 66 meV/Fe and a magnetic moment of 1.77 $\mu_B$ on each Fe atom.

Finally, we discuss electron-doping mechanism. For FeSe on STO (001) surface, FeSe is heavily electron-doped, which has been argued due to oxygen vacancies. However, we find that the doping effects from oxygen vacancies on STO (110) surface are different from those on STO (100) surface. Without a strong binding between the FeSe layer and the substrate, the oxygen vacancies donate little electron carriers to the FeSe layer. Therefore, the observed heavy doping concentration may suggest that there is a strong coupling between the FeSe layer and the substrate \cite{Zhou2015,Zhang2015}. In the meanwhile, this understanding also explains why the doping concentration observed in the FeSe on STO (110) surface is relatively lower than those observed in the FeSe on STO (001) surface\cite{Zhou2015,Zhang2015}.

To simulate the effect of oxygen vacancies, we carry out calculations for the system with an oxygen vacancy in the $1\times 2$ unitcell for the $3\times1$ reconstruction. We consider one and two oxygen vacancies out of fourteen oxygen atoms in the top surface layer, corresponding to 7\% and 14\% vacancy concentration, respectively. Our calculations show that without relaxation the vacancy-induced states are below the Fermi level and fully occupied , in contract to the case of STO (001) surface, where the vacancy-induced states are above the Fermi level. Fig.\ref{bandSTOFeSevac}(a) shows the band structure for the system with a vacancy in the top layer. The flat band marked by an black arrow in Fig.\ref{bandSTOFeSevac}(a) represents the vacancy-induced states. According to the band alignments shown in Fig.\ref{bandSTOFeSevac}(b), we find that the vacancy-induced states in FeSe/STO(001) surface can easily be empty by donating electrons to FeSe layers. However, vacancy-induced state in FeSe/STO(110) surface is filled therefore the effective doping in FeSe layer is very small. In the above calculation, the coupling between the FeSe and the substrate is small. If the coupling is large as argued in Ref.\onlinecite{Bang2013}, a strong binding between the FeSe and oxygen vacancies can change the above simple picture of energy levels to allow electron doping from oxygen vacancies.  However, because of the stability caused by the formation of TiO$_4$ tetrahedra on the STO (110) surface from the reconstruction,  compared to the case of STO(001) surface, the density of oxygen vacancies is expected to be lower \cite{Li2011}. This is consistent with the  observed electron doping  being lower  in monolayer FeSe on STO (110) surface than on STO (001) surface\cite{Zhang2015}.

\section{Discussion and Summary} \label{S3}

\begin{figure}[t]
\centerline{\includegraphics[height=4 cm]{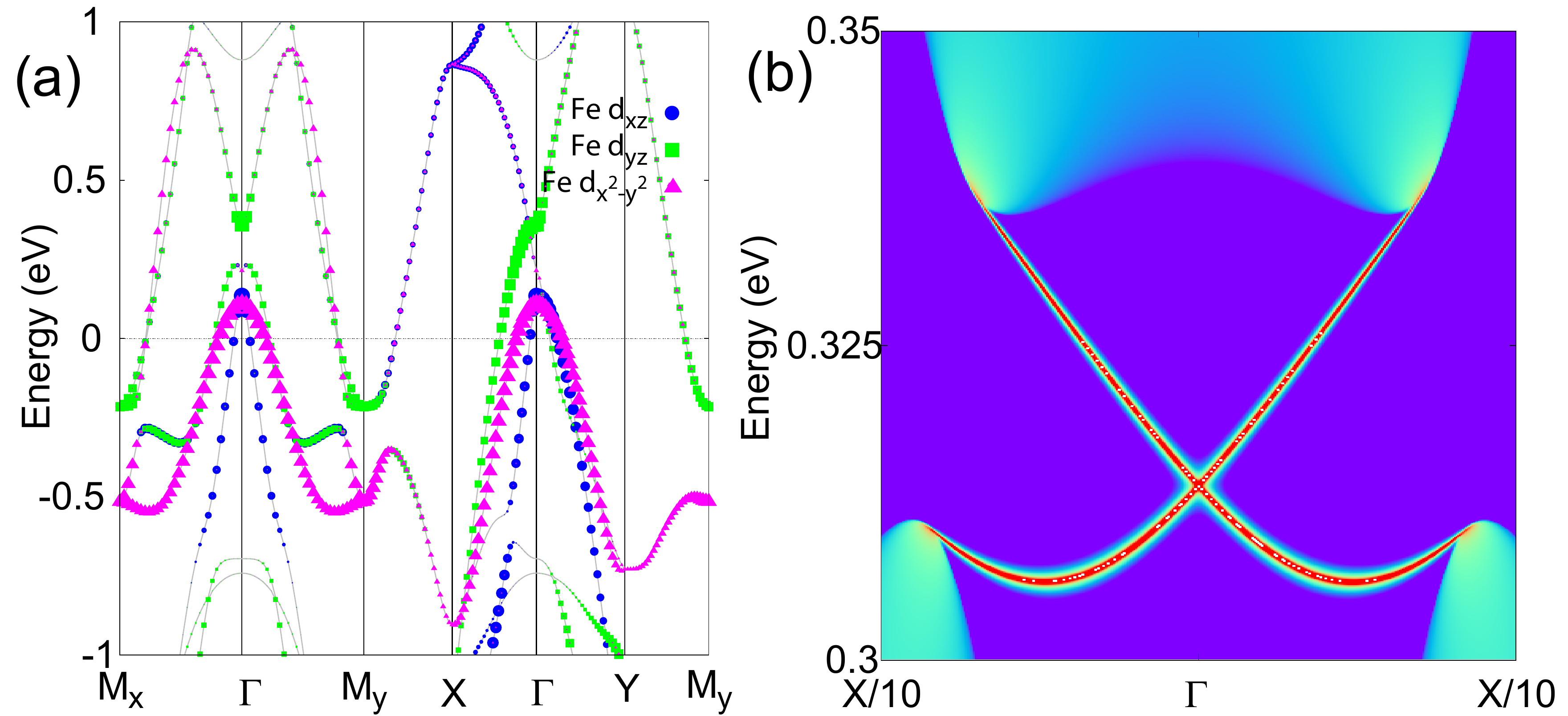}} \caption{(color online). (a) Band structure and Fermi surfaces for monolayer FeSe with $a=3.8$\AA~and 5\% compression along $b$ direction. (b) Energy and momentum dependence of the LDOS for the monolayer FeTe with strain on the [010] edge. The higher LDOS is represented by brighter color.
 \label{edgea38} }
\end{figure}

In FeSe systems, the $d_{x^2-y^2}$ band (in two Fe lattice) with odd parity at $\Gamma$ point is close to $d_{xz/yz}$ band with even parity\cite{Wu2014}.  With compressive strain along $y$ axis, the $d_{yz}$ band can shift up and anti-cross with $d_{x^2-y^2}$ band, resulting in a band inversion at $\Gamma$ point. With experimental lattice parameters and strain, $d_{x^2-y^2}$ band and $d_{yz}$ band are very close to each other. The band inversion can happen in the lattice with smaller constants or larger Se height. We plot the band structure of strained FeSe with $a=3.8$\AA~ and $b=0.95a$ in Fig.\ref{edgea38}, where band inversion has taken place. With spin-orbit coupling, the system is topologically nontrivial. The nontrivial edge states are shown in Fig.\ref{edgea38}(b). As the anion height increases linearly with the concentration $x$ in FeTe$_{1-x}$Se$_x$, we expect that this nontrivial topology can be easily realized in Fe(Te,Se) on STO (110) surface. Although the Dirac cone is far above the Fermi level in the calculations, the $d_{xz/yz}$ bands sink below the Fermi level in experiment thus these nontrivial edge states may be near the Fermi level in real materials, which could be used to realize Majorana modes below T$_c$.

From the results presented above, we find that the C$_4$ breaking strain has great influence on the hole pockets at $\Gamma$ point but little on the electron pockets except an elongation in the direction of the strain. It may explain that the superconducting gap is close to that of FeSe on STO (001) surface because there are only electron pockets in both cases. Due to the $C_4$ symmetry breaking, $s$-wave and $d$-wave pairing states can  be mixed to  result in an anisotropic gap. However, the observed gap in experiment is rather isotropic. It indicates that the $d$-wave component of the pairing state is very small and strongly unfavored, which is consistent with the pairing selection rule\cite{Hu2012,Hu2015}.

 In iron based superconductors, different superconducting mechanisms based on magnetic or orbital fluctuations have been proposed\cite{Mazin2008,Seo2008,Kontani2010}. With the strain on FeSe, both CAFM and CBAFM magnetic orders weaken but the energy difference between them decrease, suggesting   stronger magnetic frustration. This frustration suppresses long-range magnetic orders to promote  superconductivity, thus the superconducting gap for FeSe on STO (110) surface is close to that for FeSe on STO (001) surface. The electron phonon is considered to give an enhancement of superconductivity for FeSe on STO (001) surface\cite{Lee2014}. Nevertheless, the phonon frequency should strongly depend on the surfaces of the substrate. The STO (110) surface after complicated reconstructions is very different from the STO (001) surface, resulting different active phonon frequencies. However, the superconducting gaps in both cases are close, which may indicate that the electron-phonon coupling is not the primary interaction boosting superconductivity. As this anisotropic strain can suppress the long-range magnetic orders which compete with superconductivity\cite{Zhang2015}, we expect that T$_c$ in FeSe on STO (110) surface may be further increased with higher electron doping which can be achieved, for example,  through potassium deposition.

In summary, we investigate the electronic and magnetic properties of monolayer FeSe on STO (110) surface.  With compressive strain along $[1\bar{1}0]$ direction from the substrate, the monolayer FeSe has a staggered Fe lattice composed of two inequivalent Fe sublattices with a large height difference along the out-plane direction. The electron pockets and two hole pockets elongate along the direction of the strain but the remaining hole pocket elongates along the orthogonal direction. The direction of elongation for electron pockets is consistent with experimental observations. The anisotropic strain can weaken the long-range antiferromagnetic orders to enhance magnetic frustration. A low concentration of oxygen vacancies in STO (110) surface in the top layers can induce only very small electron doping to the FeSe layer. Furthermore, the strain can induce a band inversion at $\Gamma$ point and drives monolayer FeSe into a topologically nontrivial phase. The absence of strong superconducting suppression on the staggered Fe lattice suggests that the pairing configuration in real space is primarily formed within each sublattice.

\section{Acknowledgments}
We acknowledge P. Zhang and H. Ding for extremely useful discussions. The work is supported by the Ministry of Science and Technology of China 973 program(Grant No. 2015CB921300), National Science Foundation of China (Grant No. NSFC-1190020, 11334012, 11175248 and 11104339), and   the Strategic Priority Research Program of  CAS (Grant No. XDB07000000).

\end{document}